\documentclass[pra,twocolumn,showpacs,preprintnumbers,amsmath,amssymb]{revtex4-1}
\usepackage{graphicx}
\usepackage{dcolumn}
\usepackage{bm}
\usepackage{amsmath}
\usepackage{url}
\usepackage{color}
\DeclareGraphicsExtensions{.pdf,.eps,.png,.jpg,.mps} 

\begin{document}
\preprint{}

\title{Tight-binding  model for opto-electronic properties of penta-graphene nanostructures}

\author{Sergio Bravo}
\affiliation{Departamento de F\'{i}sica, Universidad T\'{e}cnica
Federico Santa Mar\'{i}a, Casilla 110-V, Valpara\'{i}so, Chile}
\affiliation{Materials Science Factory, Instituto de Ciencia de Materiales de Madrid, Consejo Superior de Investigaciones Cient\'\i ficas, C/ Sor Juana In\'es de la Cruz 3, 28049 Madrid, Spain}

\author{J. D. Correa}
\affiliation{Facultad de Ciencias  B\'asicas, Universidad de Medell\' \i n, 
Medell\' \i n, Colombia}

\author{Leonor Chico}
\email{leonor.chico@icmm.csic.es}
\affiliation{Materials Science Factory, Instituto de Ciencia de Materiales de Madrid, Consejo Superior de Investigaciones Cient\'\i ficas, C/ Sor Juana In\'es de la Cruz 3, 28049 Madrid, Spain}

\author{M. Pacheco}
\email{monica.pacheco@usm.cl}
\affiliation{Departamento de F\'{i}sica, Universidad T\'{e}cnica
Federico Santa Mar\'{i}a, Casilla 110-V, Valpara\'{i}so, Chile}

\date{\today}

\begin{abstract}

We present a tight-binding parametrization for penta-graphene that correctly describes its electronic band structure and linear optical response. The set of parameters is validated by comparing to {\it ab-initio} density functional theory calculations for single-layer penta-graphene, 
 showing a very good global agreement. We apply  this parameterization  to penta-graphene nanoribbons, achieving an adequate description of quantum-size effects.  Additionally, a symmetry-based analysis of the energy band structure and the optical transitions involved in the absorption spectra is introduced, allowing for the interpretation of the optoelectronic features of these systems.
\end{abstract}

\maketitle

\section{Introduction} 

The discovery of graphene has stimulated the quest for novel two-dimensional (2D) materials, resulting in the experimental synthesis and the theoretical prediction of various layered systems with diverse properties \cite{Novoselov_2005a,Xu2013,Balendhran2015}.
  Besides elemental analogs of graphene, such as silicene, germanene or stanene, hexagonal 2D crystals such as boron nitride or transition-metal dichalcogenides are the focus of intense research, both applied and fundamental \cite{Molle2017}. Many of these materials can be obtained by mechanical exfoliation of a three-dimensional crystal composed of weakly interacting layers coupled by van der Waals forces, like graphene itself. In fact, crystals with weakly bonded layers are being examined as a source of new bidimensional materials. Beyond mechanical methods, chemical exfoliation techniques have also been applied to covalently bonded layers to produce such 2D materials \cite{ Riedl2009,Naguib2011}. 

Inspired by such techniques, it has been theoretically proposed that penta-graphene (PG), a new 2D carbon allotrope, can be obtained from T12-carbon by breaking the covalent bonds between layers \cite{Zhang2015}. 
Although PG is a metastable carbon allotrope compared to graphene, it is energetically more favorable than the icosahedral fullerene C$_{20}$ or the smallest nanotube, which have been synthesized.  So despite some claims regarding its instability \cite{Ewels2015,Avramov2015}, it is reasonable to expect that PG might be experimentally viable. 

 Penta-graphene has been predicted to possess several unique characteristics. It  is not completely planar, and it does not have a hexagonal lattice; its 2D projection is the Cairo tiling, being composed of fused pentagons. From the electronic viewpoint, it is a semiconductor with a quasi-direct bandgap \cite{Zhang2015}, 
 being attractive for optoelectronic applications. 
PG has a reduced thermal conductivity compared to graphene \cite{Xu2015,Liu2016,QianWang2016} 
and it is an auxetic material, i.e., it has a negative Poisson's ratio \cite{Zhang2015,Sun2016}. 
It has been proposed 
as an anode material in alkaline batteries \cite{Xiao2016}, 
as a metal-free catalyst for CO oxidation \cite{Krishnan2017} 
and for use in hydrogen storage systems \cite{Enriquez2016}.
Furthermore, by doping and functionalization, the mechanical, optical, and electronic properties of PG can be tuned. 
Hydrogenation and fluorination of PG have also been analyzed, with focus on the consequences in the band gap variation \cite{QuijanoBriones2016,Wu2016,Li2016}. 

As in other 2D systems, the properties of nanostructures with lower dimensions based in PG have been also explored.
For example, 
PG nanoribbons \cite{Zhang2015,Rajbanshi2016,He2017,Yuan2017}
multilayer PG \cite{Avramov2015,Yu2015}
and 
PG nanotubes \cite{Zhang2015,Avramov2015,Chen2017}, 
which might be even more stable than monolayer PG. 
Most of these works employ a first-principles approach; recently, a tight-binding (TB) model has been put forward, allowing for the obtention of the electronic bands and an analytical expression for the optical absorption \cite{Stauber2016}. 
In fact, Zhang {\it et al.} also provided a minimal tight-binding parameterization in their pioneering work, but with a limited agreement to the ab-initio bands \cite{Zhang2015}. 

The aim of this work is to present a tight-binding parameterization of penta-graphene with an emphasis in the quantitative description of its optical spectrum, valid also for wide PG nanoribbons for which ab-initio calculations are costly. 
We choose an edge termination that does not make the ribbons magnetic, in order to focus on size effects in these systems. 
While ARPES measurements provide a direct comparison to the electronic bands, optical spectra are a standard characterization tool that is crucial for the identification of semiconductor materials, hence our motivation for this approach. We obtain our parameters by a fit to the ab-initio bands, checking that the resulting parameters give a good description of the optical absorption in PG.  Additionally, we perform a symmetry analysis of the band structure and the optical spectra of these systems.

The paper is organized as follows. In section 2  we describe the lattice structure and symmetry of the system and the computational methods. In  section 3 we explain the tight-binding parameterization  and  present our results  for the electronic structure of a monolayer PG computed with this model, and compare it with the ab-initio bands. In the same  section we also show the optical absorption response of PG and penta-graphene nanoribbons calculated with the tight-binding model, along with the corresponding ab-initio result.  Finally in section 4, we finish with some conclusions.  
 
\section{Geometry and computational 
methods} 


\subsection{Penta-graphene lattice geometry}

\begin{figure}[h!]
\centering
\includegraphics[width=0.6\columnwidth,clip]{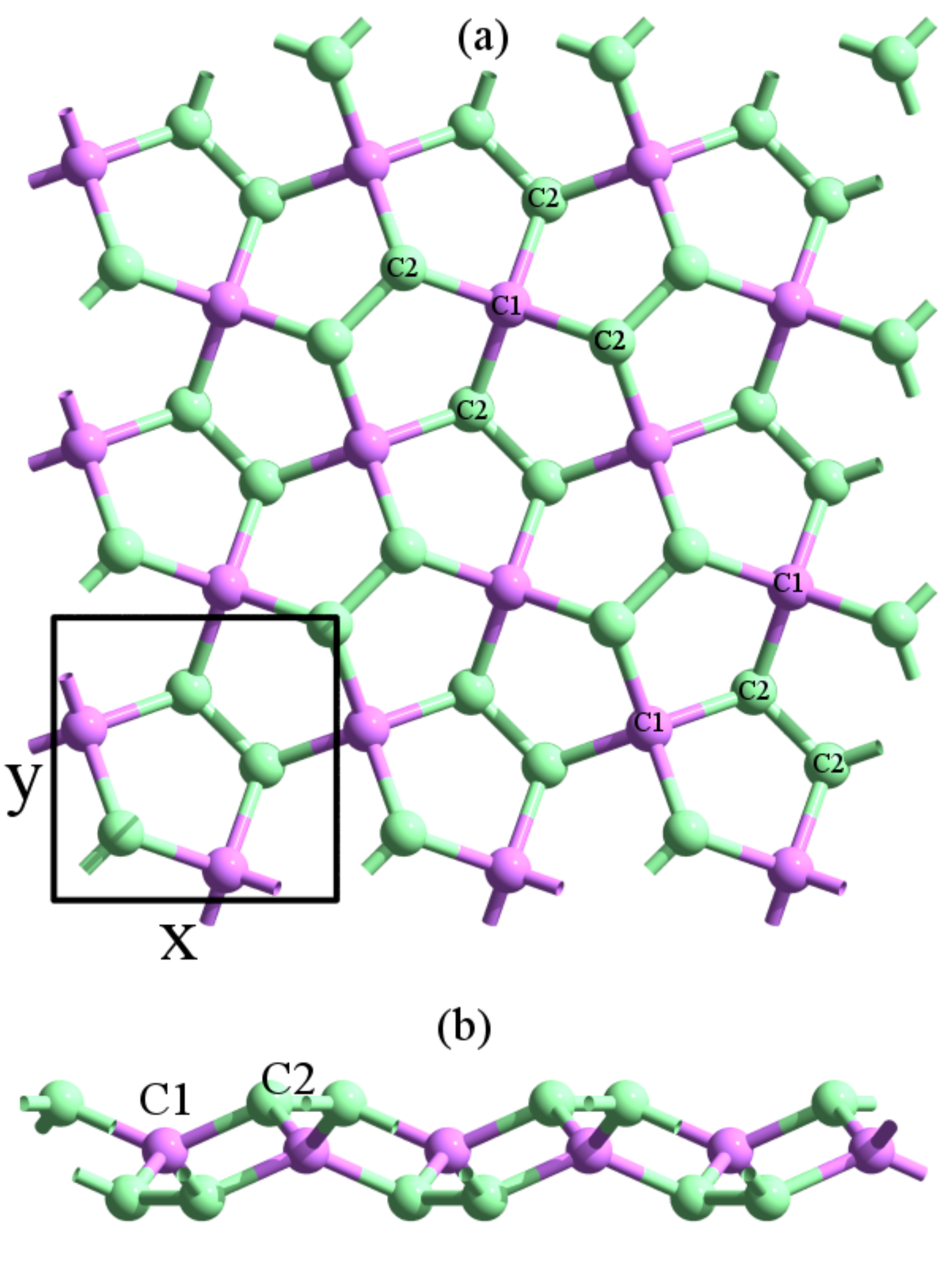}
 \caption{(a) Top and (b) side views of the PG lattice. The unit cell comprising 6 carbon atoms is enclosed in a black square. Atoms with coordination number 4 are labeled as C1, those with coordination number 3 are labeled C2. }
  \label{figlattice}
\end{figure}

As described by Zhang.~{\it et al.}~\cite{Zhang2015}, 
 penta-graphene has a 
buckled lattice structure composed by non-planar carbon pentagons, shown in Fig. \ref{figlattice}. The space group of this crystal lattice
is P$\overline{4}2_{1}$m (\#113) \cite{MillerLoveBook,Bilbao}, which
is nonsymmorphic. 
 The unit cell has 
 six carbon atoms, highlighted with a black box in Fig. \ref{figlattice} (a). The buckled lattice structure of PG can also be described as composed of three layers, see Fig. \ref{figlattice}(b). Notice that two of the atoms in the unit cell, labeled C1, have coordination 4. They belong to the central layer, whereas the other four C2 atoms have coordination 3 and form the outer layers of PG. The difference in coordination number is obviously related to a different hybridization: C1 atoms have a $sp^3$ character, whereas C2 atoms are more $sp^2$-like. 

\subsection{Computational methods}
Since the tight-binding model requires the adjustment of empirical parameters, it is necessary to resort to 
 ab-initio results in order to
fit their values, given that for the time being, there are no experimental data available. 
The optical absorption is calculated within the electric dipole approximation in both approaches. 

\subsubsection{Ab-initio approach} 

We employ the Density Functional Theory (DFT) approach, using the SIESTA  ab-initio code \cite{Soler2002}, to calculate the opto-electronic properties of monolayer PG and penta-graphene nanoribbons (PGNR). In particular, we use for the exchange-correlation functional the generalized gradient approximation (GGA) of Perdew-Burke-Ernzerhof \cite{PBE1996} 
instead of more expensive hybrid functionals \cite{Zhang2015}, because the observed general trends in the electronic structure remain almost unaltered in both schemes, and only the magnitude of the band gap is changed \cite{Zhang2015,Stauber2016}. 
We use a double-$\zeta$ plus polarization basis set and norm-conserving pseudopotentials. The mesh cutoff was set to 150 Ry and the energy shift to 0.07 eV. The Brillouin zone was sampled with a $15 \times 15 \times 1$ Monkhorst-Pack grid for PG and a $10 \times 1 \times 1$ Monkhorst-Pack grid for PGNRs. A conjugate gradient self-consistent procedure was used to relax all structures with a maximal force tolerance per atom of 0.04 eV/${\rm \AA}$. These sets of parameters assured a good energy convergence. In the case of PGNR, edges were passivated with hydrogen in order to saturate the dangling bonds.

For the optical absorption, we use a $101\times 101\times 1$ k-point grid with a broadening of 0.1 eV for PG and a $101\times 1\times 1$ k-point grid with a 0.06 eV broadening for the  PGNRs. We also assume that the electromagnetic (EM) radiation is incident perpendicularly to the PG sheet, i.e., with the electric field $\mathbf{E}$ polarization fixed in the $xy$ plane.

\subsubsection{Tight-binding model}

We follow the Slater-Koster (SK) approach \cite{SK1954} for orthogonal tight-binding calculations with the aim of providing the simplest model with a good description of the electronic and optical properties. 
Our first concern is the orbital basis choice. 
Penta-graphene only has carbon atoms, 
 so we take the usual basis selection of one $s$ orbital and three $p$ orbitals per atom. 
 There are $6$ atoms in the unit cell of PG; thus our basis for the SK Hamiltonian has $24$ orbitals.
 Note that previous parameterizations 
 with fewer orbitals present a poor agreement with DFT bands; only consideration of the full $sp^3$ basis provides a reasonable accord 
 \cite{Stauber2016}. Our goal is to find a set of parameters which gives not only a good depiction of the band structure, but also of the optical properties, so nanostructures based in PG, such as nanoribbons and nanotubes could be described within this approach in a computationally affordable manner. 

Within the electric dipole approximation,
 the optical absorption coefficient is given by 
\begin{equation}
\alpha \left(  \omega\right)  =\frac{4\pi^2 e^2}{n_0c \,m^2\omega}\sum_{c,v,\mathbf{k}}|P_{cv}(\mathbf{k}
)|^{2}\delta\left(  E_{c}(\mathbf{k})-E_{v}(\mathbf{k})-\hbar\omega\right)  , 
\label{abs}
\end{equation}
where $e$ and $m$ are the charge and  mass of the electron respectively, $\omega$ is the frequency of the EM radiation, $n_0$ is the refraction index, $c$ is the velocity of light (here set to the vacuum values) and $P_{cv}(\mathbf{k})$  are the electric dipole matrix elements $P_{cv}=\left\langle c,\mathbf{k}|\mathbf{u\cdot r|}v,\mathbf{k}\right\rangle =\mathbf{u\cdot
}\left\langle c,\mathbf{k}|\mathbf{r}|v,\mathbf{k}\right\rangle$, where $\mathbf{u}$ is the polarization vector of the external electric field and $\mathbf{r}$ is the vector position operator, evaluated between eigenstates $|v,\mathbf{k}\rangle$  and $|c,\mathbf{k}\rangle$ of the valence $(v)$ and  conduction$(c)$ bands with eigenenergies  $E_{v}(\mathbf{k})$ and $E_{c}(\mathbf{k})$, respectively.

In order to 
evaluate the absorption coefficient within the 
TB approximation, we express the matrix elements of the position operator in terms of the tight-binding parameters. 
In line with the procedure given in Refs. \cite{Voon1993,Graf1995,Zarifi2009}, 
we use the following identity for the position operator matrix, 
\begin{equation}
\left\langle c,\mathbf{k}|\mathbf{r}|v,\mathbf{k}\right\rangle =\frac{1}{E_{c}(\mathbf{k})-E_{v}(\mathbf{k})
}\left\langle c,\mathbf{k}|\mathbf{[H,r]}|v,\mathbf{k}\right\rangle ,
\end{equation}
where $H$ is the unperturbed Hamiltonian of the system.  
Expanding the eigenstates of $H$ into a linear combination of atomic orbitals
$\left\vert n,\mathbf{k}\right\rangle =\sum_{a}C_{n}\left(  a\right)
\left\vert i,\mathbf{k}\right\rangle $, 
%
where $n$ is the band index, $i$ is the atomic orbital index and $C_{n}\left(i\right)  $ are the expansion coefficients, we obtain 
the following expression for the dipole matrix elements $P_{cv}$,
\begin{equation}
P_{cv}  =\mathbf{u\cdot}\sum_{i,j}\frac{C_{c}^{\ast}\left(
i\right)  C_{v}\left(  j\right)  \sum_{\mathbf{R}_{ij}%
}\mathbf{R}_{ij}e^{i\mathbf{k\cdot R}_{ij}}t_{ij}\left(  \mathbf{R}%
_{ij}\right)
 }{E_{ck}-E_{vk}} 
\end{equation}
where $\mathbf{R}_{ij}$ are 
the lattice vectors and $t_{ij}\left(\mathbf{R}_{ij}\right)  $ represent the Slater-Koster 
TB 
parameters. 
With this expression, we can compute the optical absorption coefficient shown in Eq. (\ref{abs}). 

\section{Results}

\subsection{Tight-binding parameterization }

Since we follow the Slater-Koster scheme we have to assign to the orbital integrals the corresponding parameters. 
For PG it is already known that a simple scaling of a graphene-based parameterization yields a qualitative agreement, and a fit to DFT bands is needed to improve this description \cite{Stauber2016}. 
 With this purpose we analyze the bonding structure and geometry of PG. 
As discussed in the previous Section, in the 6-atom unit cell of PG 
there are two carbon atoms with coordination 4 and $sp^3$ hybridization, i.e., with four bonds each, labeled C1, and four carbon atoms with three bonds each, labeled C2, with $sp^2$  character. This partition leads us to treat each group of atoms separately
with respect to the 
SK parameterization. 
The basic idea is that these two groups of atoms not only have different nearest-neighbor (NN) distances, but also different hybridizations.
 From Fig. \ref{figlattice} 
we see 
that the first NNs for C1 atoms are four C2 atoms. In turn, the C2 atoms 
only have one NN, a C2 atom which is in the same layer. 
Therefore, we can assign a group of first NN parameters for each group. We parameterize the C1-C2
interaction with the Slater-Koster integrals $V_{ss\sigma }^{C1C2}$,$V_{sp\sigma }^{C1C2}$,$V_{pp\sigma }^{C1C2}$,$V_{pp\pi}^{C1C2}$, and the C2-C2 interaction with 
integrals $V_{ss\sigma }^{C2C2}$,$V_{sp\sigma }^{C2C2}$,$V_{pp\sigma }^{C2C2}$,$V_{pp\pi }^{C2C2}$.
On the other hand,
for the C2 atoms we already have included up to second NNs. 
 We consider also the hoppings between a 
C2 carbon atom in one of the external planes and another C2 atom from the opposite one, which we have labeled as C2$^{\prime }$ 
to distinguish it from 
the first NN C2-C2 pair. This interaction is indeed a third NN interaction, and we can assign the corresponding
Slater-Koster integrals $V_{ss\sigma }^{C2C2^{\prime }}$,$V_{sp\sigma}^{C2C2^{\prime }}$,$V_{pp\sigma }^{C2C2^{\prime }}$,$V_{pp\pi}^{C2C2^{\prime }}$. 

This exhausts the basic interactions for our 
model. We have checked that considering the next NN for the C1 atoms, i.e., another C1-C2 coupling, does not improve appreciably our results, so in fact we 
have a geometrical cutoff that includes interactions up to distances equal or smaller that the 3rd NN interactions between C2 atoms. 

In summary, we have twelve 
SK hopping parameters with contributions up to first NN for the C1 atoms and up to third NN for the C2. 
Finally, we 
consider the onsite energies associated with each atom and orbital. Specifically, we assign four onsite energies, $E_{s}^{C1},E_{p}^{C1},E_{s}^{C2},E_{p}^{C2}$ corresponding to
the $s$ orbital and the three $p$ orbitals for each group of atoms respectively. 
This amounts to 
a total of sixteen 
 SK parameters in our model.

The fitting is customarily done with respect to the bands obtained within the DFT approach. However, we have verified that the four bands closer to the Fermi level may have an excellent agreement with the DFT bands, but without achieving a good fit with respect to the optical absorption results. We have carried out a least-square minimization procedure looking for a compromise between a good fit to the bands and an adequate description of the optical properties. 
The final values for 
the 
TB parameters 
are presented in Table \ref{latabla}.

\begin{table*}[]
\begin{ruledtabular}
\begin{tabular}{cccccccccccccccc}
$E_{s}^{C1}$         & $E_{p}^{C1}$         & $E_{s}^{C2}$         & $E_{p}^{C2}$  &  $V_{ss\sigma}^{\rm C1C2}$  & $V_{sp\sigma}^{\rm C1C2}$  & $V_{pp\sigma}^{\rm C1C2}$  & $V_{pp\pi}^{\rm C1C2}$ & $V_{ss\sigma}^{\rm C2C2}$  & $V_{sp\sigma}^{\rm C2C2}$  & $V_{pp\sigma}^{\rm C2C2}$  & $V_{pp\pi}^{\rm C2C2}$ & 
$V_{ss\sigma}^{\rm C'_{2}C2}$ & $V_{sp\sigma}^{\rm C'_{2}C2}$ & $V_{pp\sigma}^{\rm C'_{2}C2}$ & $V_{pp\pi}^{\rm C'_{2}C2}$  \\ \hline
$-2.492  $                      & 5.549                         & 12.035                        & $-2.125 $      &   -3.789                        & 0.530                         & 9.270                         & -1.050    & $7.383$                        & $-2.181$                       & 6.981                         &- 0.671                   & $-4.054   $                     & 3.809                         & 0.545                         & -0.099                     \\   
\end{tabular}
\caption{ Slater-Koster tight-binding parameters (in eV) for PG. }
\label{latabla}
\end{ruledtabular}
\end{table*}

\subsection{Electronic properties of PG } 

\begin{figure}[h!]
\centering
\includegraphics[width=0.9\columnwidth,clip]{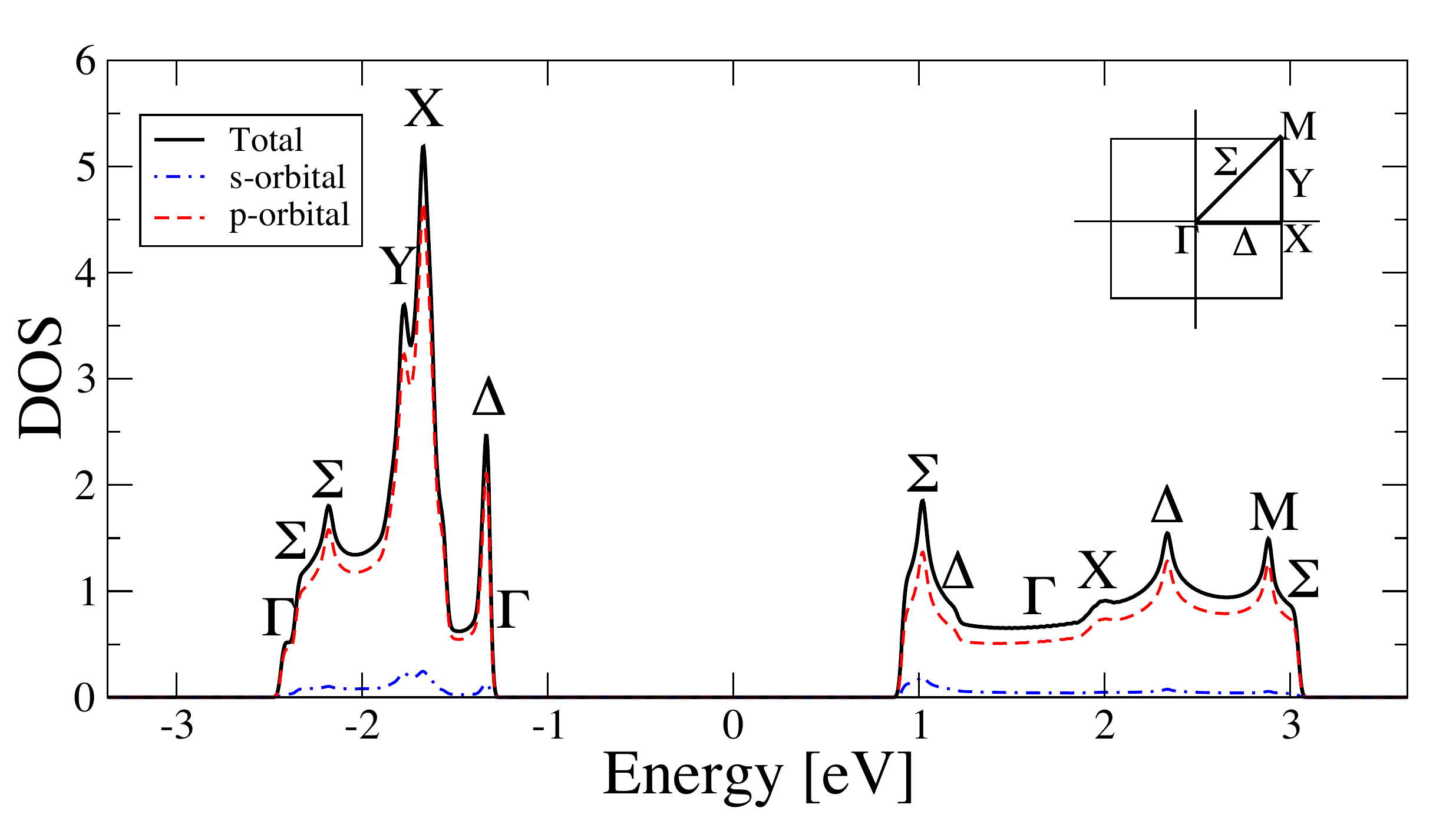}
 \caption{Penta-graphene total DOS (black solid line), $p$-orbital projected DOS (red dashed line) and $s$-orbital projected DOS (blue dashed-dotted line).}
  \label{DOSM}
\end{figure}

This election of tight-binding parameters 
can be supported with the orbital decomposition of the density of states (DOS) for monolayer PG, computed within the ab-initio approach. 
The DOS calculation was performed with SIESTA with a dedicated $200 \times 200 \times 1$ k-point grid for this particular calculation along with a energy broadening of 0.010 eV.
Fig. \ref{DOSM} shows the total DOS of penta-graphene, along with the orbital decomposition in the $s$ and $p$ orbitals. 
 In addition, we have identified the 
 peaks corresponding to high symmetry points and lines within the Brillouin zone (BZ), shown in the inset of the Figure, 
 with the aim to elucidate the symmetry of the 
BZ points 
 yielding high a density of states. 
This can serve as a 
  guide to understand the optical features of the material. 

\begin{figure}[h!]
\centering
\includegraphics[width=0.9\columnwidth,clip]{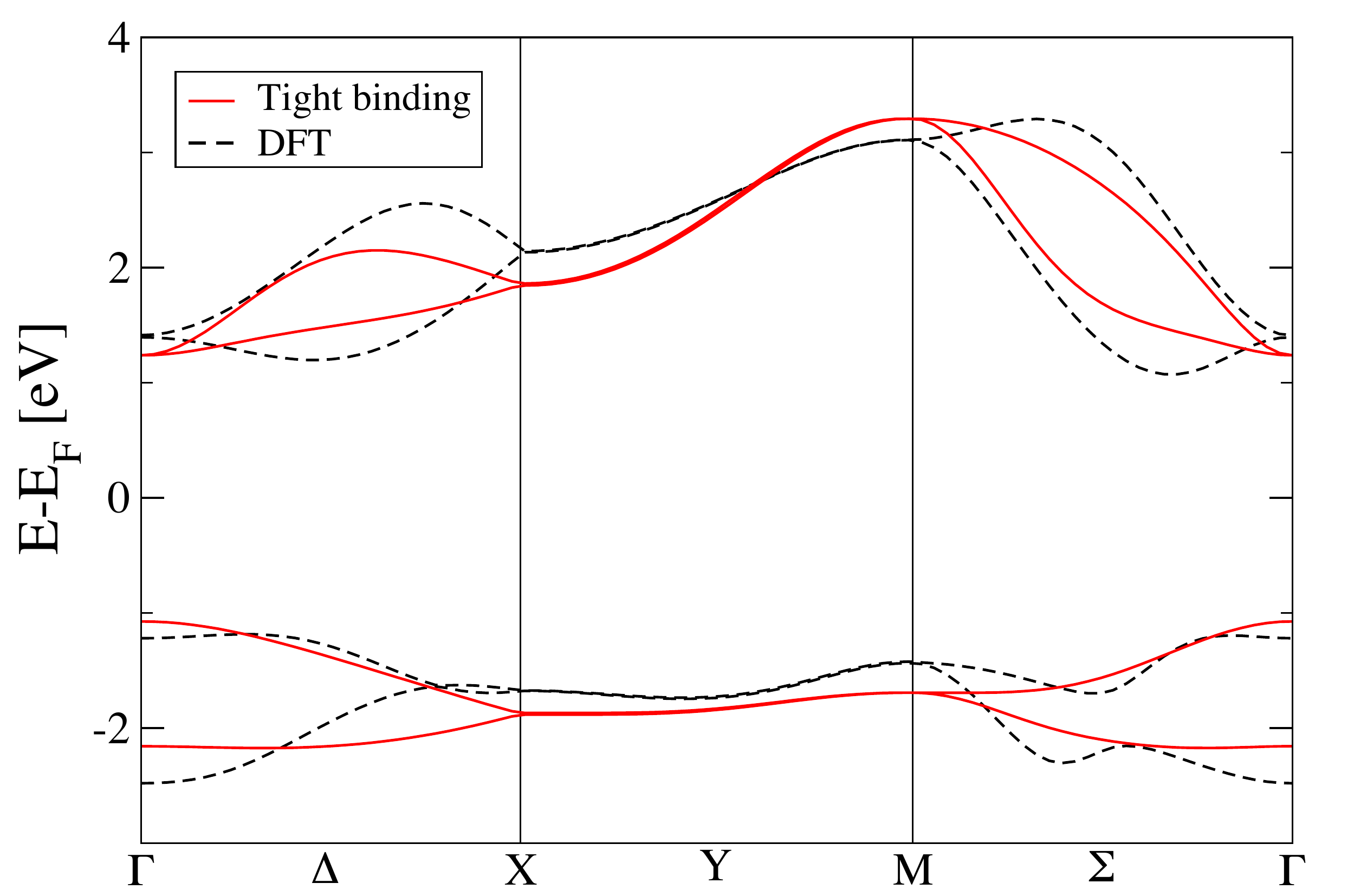}
 \caption{PG energy band structure near the Fermi level calculated with DFT (black dotted lines) and tight-binding (red solid lines).}
  \label{bandsMPG}
\end{figure}

In Fig.  \ref{bandsMPG} we present the 
TB 
band structure obtained with the parameters given in Table \ref{latabla} along with the bands calculated using the SIESTA 
code for PG.  It can be seen that there is a good overall agreement between them.  A few remarks are opportune with respect to the parameterization procedure. 
 As mentioned before,  
 it is feasible to obtain an excellent fit to the four bands of interest (two valence and two conduction bands) with the same number of parameters, but failing to reproduce other valence bands at lower energy, in such a way that the optical spectrum is not even qualitatively correct. Likewise, the quasi-direct gap can also be reproduced, but with either same consequences for the optical spectrum or requiring a much larger number of parameters. We opted for a compromise solution, maintaining the overall agreement of the bands but without losing the description of the optical properties 
while keeping the same number of parameters.
 In fact, we have checked that the quasi-direct gap does not play a substantial role in the calculation of the direct absorption. 

\subsection{Optical properties of monolayer PG}

\begin{figure}[h!]
\centering
\includegraphics[width=0.9\columnwidth,clip]{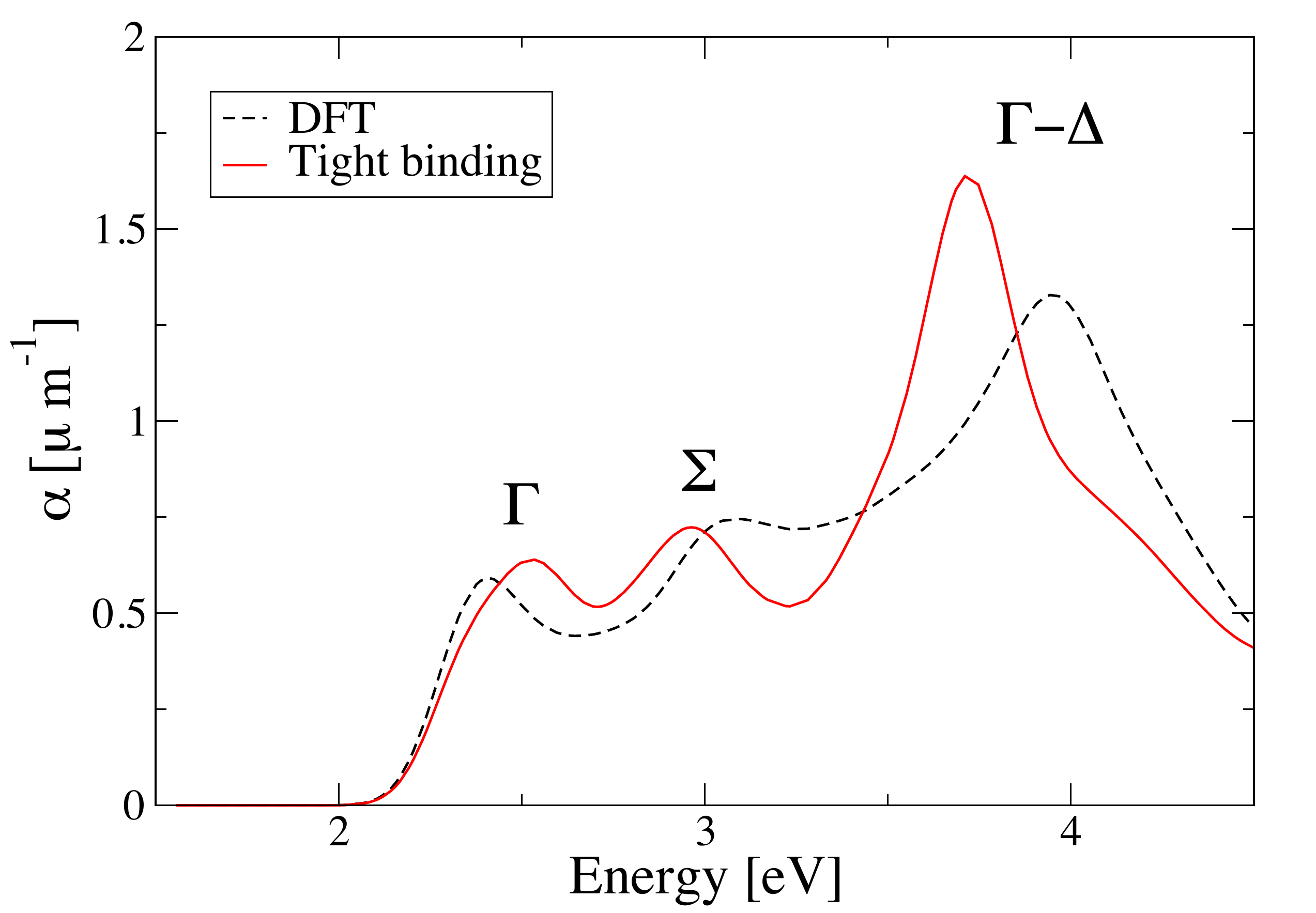}
 \caption{Optical absorption for PG calculated with tight-binding (red solid lines) and ab-initio SIESTA code (dashed black lines).}
  \label{opticsMPG}
\end{figure}

As in the case of the electronic band structure, we compare the 
 optical absorption coefficient computed within the DFT approach with that calculated with the 
 TB approximation. 
In this case we use the same $k$-space grid and 
 broadening energy in both calculations, 
TB and DFT.  
The results are shown in Fig. \ref{opticsMPG}. 
There are three marked peaks in the low energy region of the optical spectrum. We have labeled them according to the symmetry of the relevant states, 
as in the DOS plot. These peaks arise due to different transition amplitudes within the Brillouin zone. The lowest peak ($\sim$ 2.45 eV), 
is due to transitions near the band gap; it is dominated by contributions from a
region near $\Gamma$, where the conduction and valence band states have an 
energy difference $\sim$ 2.5 eV. The next peak ($\sim$ 3.1 eV) has contributions
from the $\Sigma$ line across the 
BZ, 
where many transitions are allowed due to the low symmetry. The last and higher peak in this energy 
window 
($\sim$ 3.9 eV), has two major contributions. 
One is from $\Gamma$, where symmetry allows a transition around 4 eV; the  
second is from the $\Delta$ line where 
we have several allowed transitions due to the low symmetry present in this line, similar to those from the $\Sigma$ direction 
and the peak near 3 eV. 
Additional symmetry analysis gives us another interesting property related to the optical response of PG.  Due to the group of the wave vector at the high symmetry points X and M of the BZ, we obtain a selection rule that forbids transitions from the valence to the conduction band at these two points. This is because of the different parities of the irreducible representations that are coupled by the momentum operator, giving a direct product that does not contain the invariant irreducible representation of the group \cite{DresselJorioBook}. 

Thus, the TB parameterization of Table \ref{latabla} 
reproduces these three peaks 
 in the optical absorption calculation. Only the higher energy peak shows an appreciable difference in intensity and energy comparing the DFT and TB results. 
 This is due to the fact that this peak stems from transitions in two different regions of the BZ, namely, $\Gamma$ and the $\Delta$ line; fitting these two contributions simultaneously in TB is very difficult.  
 We would like to emphasize that our parameterization 
 of the energy band structure, which uses the optical absorption as a a criterion for its validity, manages to provide a remarkably good description of 
 both features, being a solid starting point for the study of PG-based nanostructures.

\subsection{Penta-graphene nanoribbons } 

\begin{figure}[h!]
\centering
\includegraphics[width=0.7\columnwidth,clip]{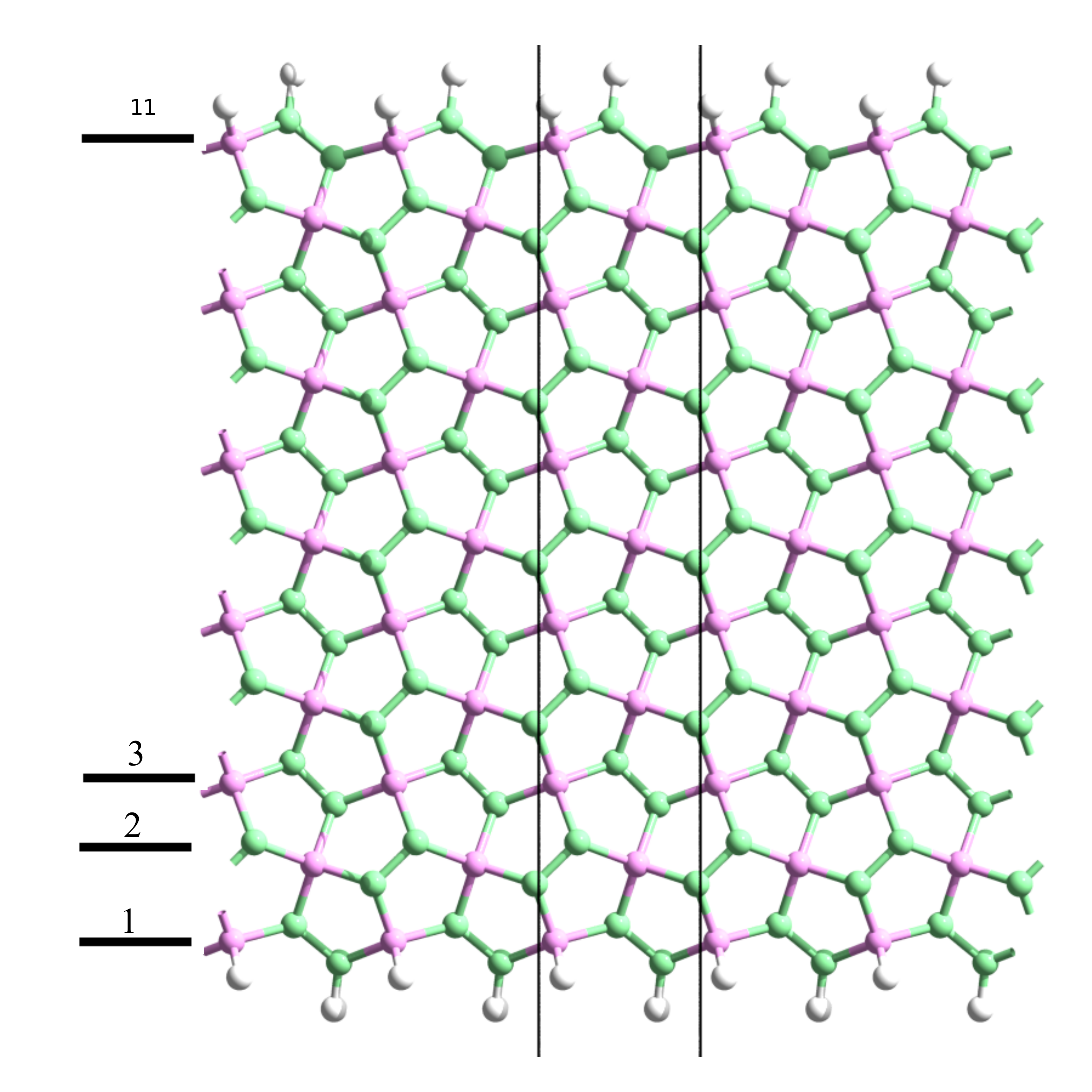}
 \caption{ 11-PGNR lattice structure. The translational unit cell is marked between two black lines. The labeling based in longitudinal chains is herein illustrated.  }
  \label{PGNRlattice}
\end{figure}

In this section we present the band structures and optical absorption spectra of a particular type of penta-graphene nanoribbons (PGNRs) 
as a means to test
 our model in nanostructured systems. In particular, we choose 
PGNRs with sawtooth-like edges \cite{Rajbanshi2016}, 
shown in Fig. \ref{PGNRlattice}. 
The reason for this choice is that PGNRs with such edges are not magnetic \cite{He2017,Yuan2017}, so we can concentrate in the validity of the tight-binding parameterization focusing on size effects. 
PGNRs are labeled with the number of longitudinal chains
across its width. 
For example,  Fig. \ref{PGNRlattice} 
depicts a 11-PGNR. 
Obviously, the symmetry of PGNRs is reduced with respect to PG.    
Since nanoribbons have translational symmetry in only one 
direction, 
we have to resort to the so called rod groups to describe their symmetry. 
The PGNRs studied in this work  
belong to the rod group labeled $P112_{1}$ \cite{MillerLoveBook,DamnMiloBook}. 
It has the identity transformation plus a $C_{2}$ rotation around the periodic axis of the ribbon combined with a glide plane translation by $1/2 a$, where $a$ is the lattice constant vector in the 
direction with translational symmetry.

\subsubsection{Band structure of PGNRs} 

Ultranarrow nanoribbons present strong lattice relaxation effects, so 
we focus on wider ribbons, for which such effects are not so important and can be potentially described without taking into account such relaxation. Furthermore, for wider ribbons the 
TB approach is more advantageous. 
 The translational unit cell employed for the 
 calculations is marked with two black lines in Fig. \ref{PGNRlattice}. 

\begin{figure}[h!]
\centering
\includegraphics[width=0.9\columnwidth,clip]{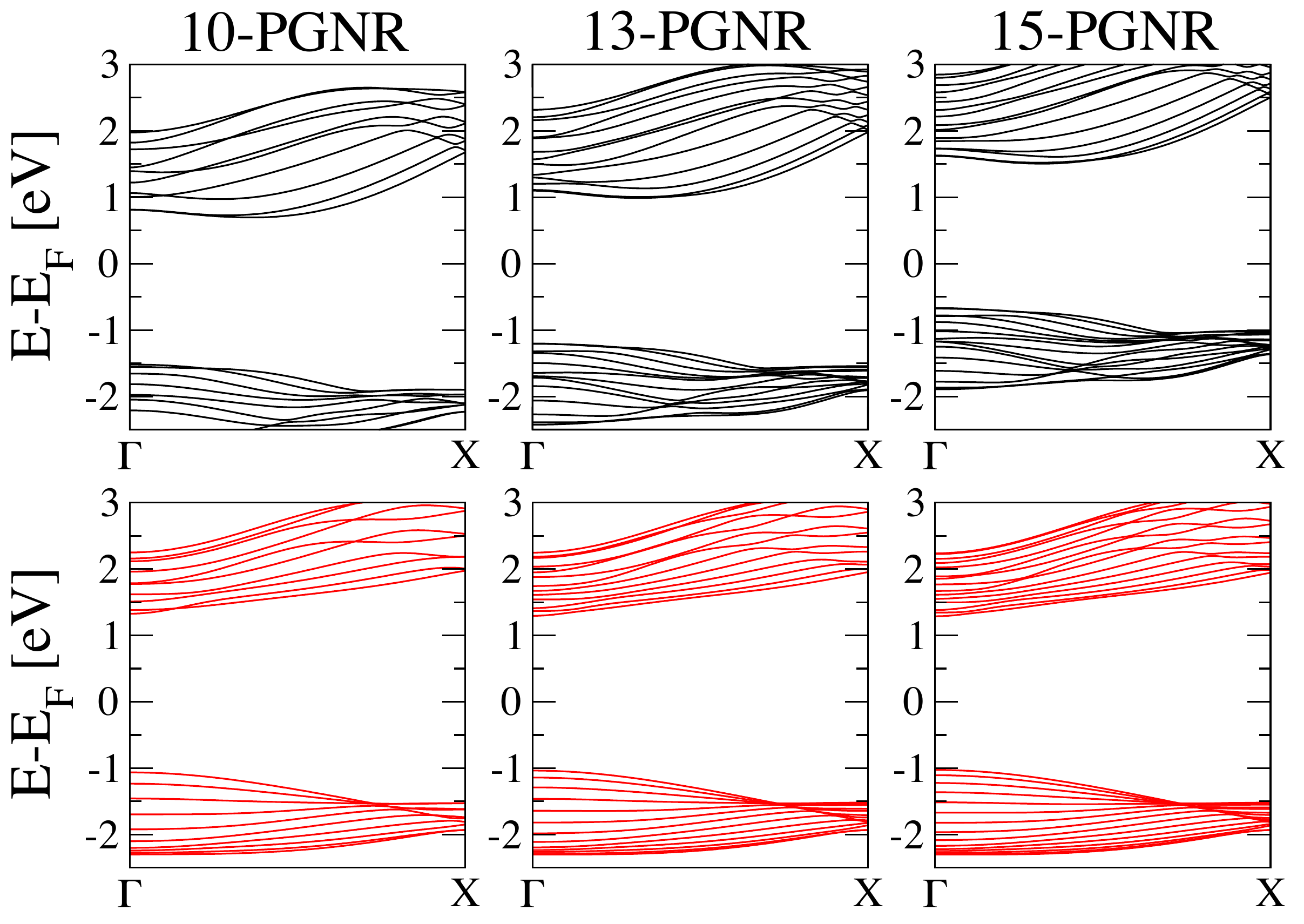}
 \caption{ Energy band structures
  for 10-PGNR, 13-PGNR and 15-PGNR, as labeled in the Figure. Top panels (black lines) are computed with SIESTA; bottom panels (red lines) are the TB results. 
   }
  \label{PGNRbands}
\end{figure}

We use a hard wall boundary condition for TB and the 
parameterization of Table \ref{latabla}.  
Fig. \ref{PGNRbands} shows the band structures calculated with SIESTA and 
TB for three particular nanoribbons, namely, 10-PGNR, 13-PGNR and 15-PGNR, respectively.
The band structures obtained by both, the ab-initio and the TB method, show a good qualitative agreement.
Valence subbands lack some of the fine details concerning some accidental degeneracies that occur 
along the $\Gamma$X  line. Most remarkably, the low-energy conduction subbands lose the indirect minima appearing in the afore mentioned 
  $\Gamma$X line. These differences can be easily understood, 
  since the band structure of the bulk PG did not reproduce the quasi-direct gap, related to these features in the nanoribbons. 
On the other hand, 
since the ribbons have a lower symmetry with respect to the bulk structure, we expect some decrease in the degeneracy at high symmetry points in the Brillouin zone. This can be seen at $\Gamma$ and X, where subbands tend to avoid degeneracy in contrast to the case of the bulk.  This is observed in both calculations, DFT and 
TB. 

\subsubsection{Optical absorption of PGNRs}

\begin{figure}[h!]
\centering
\includegraphics[width=1.0\columnwidth,clip]{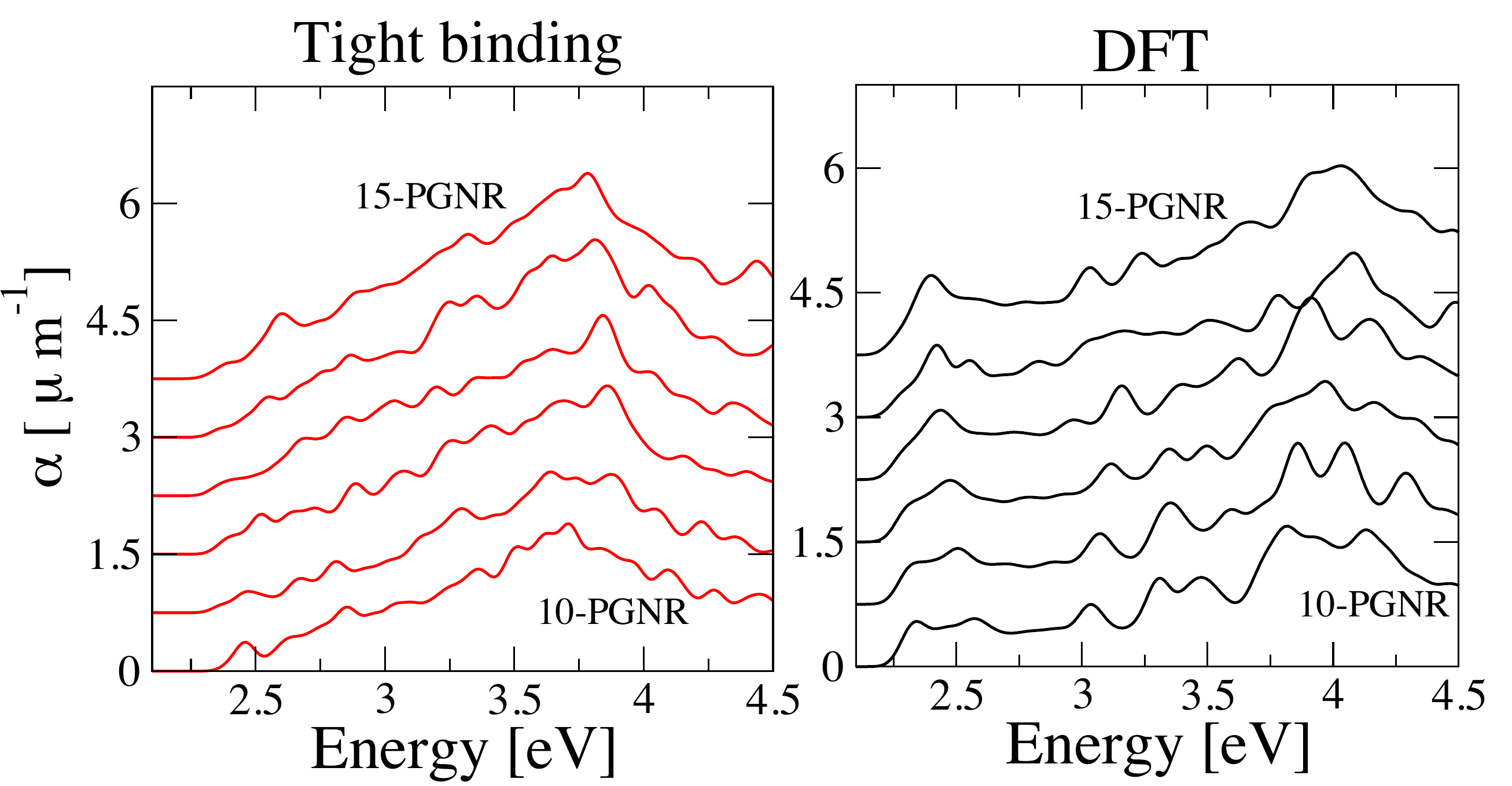}
 \caption{Optical absorption for 10- to 15-PGNRs calculated with (a) TB; and (b) DFT (SIESTA). The bottom and the top curves in each panel are labeled; each 
 result for increasing width is shifted a fixed amount. }
  \label{opticsPGNR}
\end{figure}

We have additionally computed the  
optical absorption for different widths of the nanoribbons. 
In this case we consider that the electric field of the EM radiation oscillates along the nanoribbon axis. 
In order to make a comparison between both approaches and test the
TB parameterization, we have also employed a first-principles method.  
We use the same external electric field configuration and $k$-space grid in the TB calculation as in the DFT, but with a smaller broadening of 0.03 eV. 
The optical absorption is also computed employing Eq. (\ref{abs}). 
The included nanoribbon subbands are those stemming from the bulk bands
considered for the monolayer optical absorption. 

Results for optical absorption spectra in both approaches are presented in Fig. \ref{opticsPGNR}. 
Some remarks can be noted. Taking the DFT calculation as a reference, we observe that  a double-peak structure is present in the low-energy region of the spectrum (near the energy of the band gap).
This can be understood by resorting at  the subband
 structure at $\Gamma$, which has several allowed contributions in this energy range, yielding 
  a multi-peak spectrum that stems from the vicinity of the high-symmetry 
$\Gamma$ point. As
the width of the ribbon increases, 
 the lower peak in this region starts to lose 
 weight, 
 and 
 on the
  contrary, the higher-energy peak in this pair gains
   intensity, showing a clear tendency to form the peak near the band edge energy 
also observed in the bulk. 
   This evolution of the low-energy peaks is also reproduced in a qualitative way in the TB calculation, with 
 diverse values of the   intensities. Also, the upper peak of this pair
  is located in a higher position in the tight-binding spectra with respect to the DFT case. This is related to the shift to higher energies of the $\Gamma$ peak observed in the bulk:
  due to the compromise between the fit to optical and band structure results, 
  the TB gap at $\Gamma$ is slightly wider than the first-principles gap, which results in a similar peak shift in the nanoribbons. 
 
 The same evolution with increasing ribbon width is observed for the 3.9 eV peak in the bulk monolayer DFT calculation, appearing at lower energy in the tight-binding case. 
 The spectra of the ribbons evolve leading to the appearance of this peak in wider ribbons. 
   However, there is 
   a difference with  respect to monolayer PG: 
    because of the lower symmetry of the ribbon, transitions around X are not forbidden as they are for the bulk.
    Therefore, much of the intensity of the peak 
 is due to transitions around this symmetry point.  There are also some minor contributions
     from the $\Sigma$ line (near X) and $\Gamma$. 
     In the TB calculations we see that the peak appears at a lower energy 
     with respect to the DFT result; this is also related to the underestimation of the gap at X  
     in the TB energy band structure.
The energy range between these two peaks (2.5 and 3.9 eV in DFT)  presents many smaller peaks due to quantum size effects in both calculations, SIESTA and TB. In this energy range there are contributions from states 
originating in the 
$\Delta$ line of the bulk Brillouin zone. 
 Similarly to the other two main bulk peaks, for larger widths of the nanoribbons 
  the middle peak labeled $\Sigma$ in the bulk also appears in them, thus converging to the 2D system.

\section{Conclusions}

We have presented a tight-binding parameterization for penta-graphene that provides a very good description of the opto-electronics properties of this material, as the comparison of the tight-binding calculated magnitudes to first-principles results shows. 
Our choice of parameters was guided by the existence of two types of hybridization in PG: we assigned different parameters to atoms with different hybridizations, and set a geometric cutoff corresponding to third-nearest neighbor interactions for the C2 atoms. The validity of the basis and parameterization was substantiated by the orbital-resolved DFT calculated density of states of PG and by the agreement of the energy bands and optical spectrum calculated within the TB and the DFT approaches, respectively.
This parameterization was also employed to model PG nanoribbons with non-magnetic edges, achieving a good description of the quantum-size effects and the recovery of bulk features with increasing widths.
We additionally performed a symmetry analysis of the bands, identifying the space group structure of PG and elucidating the contributions of distinct states to the prominent peaks of the optical spectra. 
Our parameterization can be of interest to model further physical properties of pentagraphene-based nanostructures, for which a first-principles approach is computationally unaffordable. 

\acknowledgments
This work has been partially supported by the Spanish MINECO under 
Grant No. FIS2015-64654-P, CSIC i-coop Grant No. ICOOPA-20150, Chilean FONDECYT Grant No.1151316 and Chilean CONICYT PhD scholarship No. 21150492.
S.B., L.C. and M.P. thank the Universidad de Medell\'{i}n for the warm hospitality. 
J.D.C. thanks the Laboratorio de Simulaci\'on y Computaci\'on Cient\'{i}fica at Universidad de Medell\'{i}n for  computational time.

%

\end{document}